\documentclass{article}

\usepackage{PRIMEarxiv}

\usepackage[utf8]{inputenc} 
\usepackage[T1]{fontenc}    
\usepackage{hyperref}       
\usepackage{url}            
\usepackage{booktabs}       
\usepackage{amsfonts}       
\usepackage{nicefrac}       
\usepackage{microtype}      
\usepackage{lipsum}
\usepackage{fancyhdr}       
\usepackage{graphicx}       
\usepackage{wrapfig}
\graphicspath{{figures/}}     
\usepackage[nameinlink]{cleveref}

\title{Toward Cyclic A.I. Modelling of Self-Regulated Learning: A Case
Study with E-Learning Trace Data
}

\author{
  Andrew Schwabe \\
  University of St. Andrews \\
  St. Andrews, Scotland\\
  \texttt{as613@st-andrews.ac.uk} \\
   \And
  Özgür Akgün \\
  University of St. Andrews \\
  St. Andrews, Scotland\\
  \texttt{ozgur.akgun@st-andrews.ac.uk} \\
   \And
  Ella Haig \\
  University of Portsmouth \\
  Portsmouth, United Kingdom\\
  \texttt{ella.haig@port.ac.uk} \\
}

\begin{document}
\maketitle

\begin{abstract}
Many e-learning platforms assert their ability or potential to improve students' self-regulated learning (SRL), however the cyclical and undirected nature of SRL theoretical models represent significant challenges for representation within contemporary machine learning frameworks.  We apply SRL-informed features to trace data in order to advance modelling of students' SRL activities, to improve predictability and explainability regarding the causal effects of learning in an eLearning environment. We demonstrate that these features improve predictive accuracy and validate the value of further research into cyclic modelling techniques for SRL.
\end{abstract}

\keywords{Artificial Intelligence \and Machine Learning \and Self Regulated Learning }

\section{Introduction}
Online education delivery continues to be a growing trend, and is often presented with claims about optimizing or improving student’s learning and self-regulation~\cite{alonso-menciaSelfregulatedLearningMOOCs2020,mdzalliOnlineSelfRegulatedLearning2020}. Generative AI (GenAI), and particularly Large Language Models (LLMs) such as ChatGPT are increasingly being used in e-learning environments with both positive and negative effects regarding knowledge retention and self-regulation~\cite{bastaniGenerativeAICan2024,stadlerCognitiveEaseCost2024,EfficacyGenAITools,ImpactGenerativeAI2025}.

Research about self-regulated learning is considered foundational regarding measurement of a student's capability to retain taught materials~\cite{panaderoReviewSelfregulatedLearning2017}.   While MOOCs, ITSs and LMSs capture a lot of data, there are missing links between the well known causal aspects of theoretical SRL and the evaluation of learning using only an e-learning system’s trace data~\cite{dimitriLearningPulseMachine2017a}.  GenAI tools are likely a complication of this by existing as external tools that may or may not even be logged in trace data.

This paper presents background and a progress update about hypothesis, methodology and experimental results to extract and model self-regulated learning patterns from students’ use of the Moodle LMS trace data~\cite{menzeInteractionReadingAssessment2022}.  The dataset is from a published study designed to identify gaps in learning based on cluster analysis of user behavior within Moodle LMS.  We utilize this data as part of ongoing research to propose new methods of modelling the cyclic nature of self-regulated learning in efforts to provide causal explanations that connect back to cause-effect learnings from well known SRL studies.  The cyclical representation of SRL theory models is an aspect commonly missing from modern machine learning applications in e-learning and SRL~\cite{franciscoPerusallsMachineLearning2021,panaderoHowStudentsSelfregulate2014}.

Findings present that more research and methods are needed to accurately model the cyclical reinforcement effect of self-regulation activities within e-learning environments.  By engineering data features that capture SRL behaviours, results show that this method can significantly improve academic outcome predictability.

\section{Background and Related Work}

Self-Regulated Learning (SRL) is an area of educational psychology research about activities and influential forces that affect and support students' developmental abilities to learn~\cite{nihr_wpSchoolbasedSelfregulationInterventions2018}. Several prominent theories have been presented, studied and applied by leading experts such as Zimmerman, Boekarts, Pintrich, Winne and Hadwin~\cite{panaderoReviewSelfregulatedLearning2017}.  A foundational concept expanded on by many of these models is that of a cyclical feedback loop, where iterative behaviours and choices reinforce and affect future SRL strategy and activities.  The Zimmerman cyclical model is often cited as the basis for cyclical self regulation, defining activities that are classified into three main phases:  \textbf{Forethought}, a planning phase, A \textbf{Performance} phase, where students apply discipline to follow their intended plans, and \textbf{Self-reflection} phase, where students evaluate the results and effectiveness of their planning and performance~\cite{boekaertsSelfregulatedLearningNew1997}.  Self-reflection activities then affect subsequent iterations of forethought and performance activities.  Insights from this repetitive process, both positive and negative, affect future planning, but also affect motivation and emotional perception of the learning task and perceived value of the work effort~\cite{afariGlobalSelfEsteemSelfEfficacy2012,raaijmakersEffectsPerformanceFeedback2017a}.

\begin{table*}[b]
\centering
\begin{tabular}{|l|l|} 
\hline
 Feature & Description \\
\hline 
 \textit{reading\_sessions} & Count of how many distinct reading sessions for the user. \\
\hline 
 \textit{num\_reading\_breaks} & Count of how many breaks were taken during reading. \\
\hline 
 \textit{quiz\_time\_mins} & Elapsed minutes from start to finish of the quiz. \\
\hline
 \textit{quiz\_fails} & How many times a quiz was failed by the student. \\
\hline
 \textit{quiz\_attempts} & How many times a quiz was started by the student. \\
\hline
\end{tabular}
\caption{Baseline Dataset Features\label{table1}}
\end{table*}

Prior to digital classroom technology, offline measures, such as surveys, helped inform and prove improved self-regulation when scaffolds or interventions are introduced to students in traditional learning environments~\cite{arakaResearchTrendsMeasurement2020,magnoValidatingAcademicSelfregulated2011a,rothAssessingSelfregulatedLearning2016}.  Many studies about academic performance and student self-regulation related to e-learning, however, rely on online measures, such as trace (log) data, and multimodal sensor data, or mixed method approaches, combining log data with surveys~\cite{dimitriLearningPulseMachine2017a,jansenMixedMethodApproach2020,moreno-marcosTemporalAnalysisDropout2020,winneSupportingSelfRegulatedLearning}.

These methods, even when combined using mixed method approaches, result in statistics biased toward the e-learning system’s delivery design and logging structure, but could benefit from incorporating metrics based on SRL strategies~\cite{moreno-marcosTemporalAnalysisDropout2020a}.  Analysis of trace data, such as in gStudy, finds statistical effects of learning activities within the context of the e-learning system itself~\cite{azevedoLessonsLearnedFuture2022,azevedoAnalyzingMultimodalMultichannel2019}.  

Studies have shown that trace data can be used effectively to detect students that are less likely to succeed, and by prescribing training as an intervention, have been able to demonstrate effectiveness of human intervention~\cite{coglianoSelfregulatedLearningAnalytics2022}. In addition, use of temporal behaviors has been shown to help predict near-term student learning outcomes~\cite{plumleyCodesigningEnduringLearning2024}.  

Although machine learning has been used in these examples to positive predictive effect, no connection or mapping is made to  SRL theory phases or inter-phase causation.  This leaves a gap in true understanding about how or why an e-learning system’s design or activities explain or demonstrate causal effect for increased self-regulation~\cite{liCognitiveEngagementSelfregulated2022,molenaarMeasuringSelfregulatedLearning2023}.

Common methods for cleaning and modelling SRL trace data result in a loss of cyclical knowledge in the data.  Common sample-based techniques for machine learning require data to fit a one-sample per row design, which encourages a bias used in analysis of the data~\cite{alyahyanPredictingAcademicSuccess2020,osakweReinforcementLearningAutomatic2023}. While some models exist for catching recurrent data trends in time series, reinforcement learning and recurrent neural networks, there are inherent problems with modelling SRL theory cyclical patterns that do not adhere to a standard directed graph approach.

\section{Case Study}

The data for this study comes from a previously published study of an online BSc course for "Operating Systems and Computer Networks." The study examined patterns and insights between graduate students’ reading activity, behavior and quizzes.  The study was implemented using Moodle, where 142 students consented and successfully recorded related behavior and quiz data.  Ages of participants ranged from 19-65 years old, and the quizzes were combinations of self-assessment questions and multiple choice answers. A unique aspect to this study is additional software tooling that tracks a student’s scrolling behaviour on specific pages, so that patterns of start, stop, restart, and reading breaks (longer than a threshold) can be quantified behaviourally and used in cluster analysis~\cite{menzeInteractionReadingAssessment2022}.  This source study was selected because of the availability of raw data, because it included suitable behavioral data, and because it was already published and peer-reviewed. In the future, we plan to conduct our own study and collect our own data.

The additional software was used to collect page scroll activity (trace data) while students read the content pages associated with each part of the course and expanded on the level of tracking provided by the LMS.  Metrics include time stamps, page object IDs, and the page scrollY from the DOM which allows tracking of how far up or down the student was scrolling. 

The scrolling trace data is then used to analyse students’ reading sessions so that the number of unique reading sessions may be quantified (defined by a re-start of reading from the top of the page).  In addition, a threshold is used to quantify reading “breaks”, defined by exceeding a threshold of time within the scrolling trace data.

\subsection{Difficulty in Modelling Cyclic SRL Theory}
SRL Theory models have inherent cyclic and undirected characteristics.  Several technologies have been evaluated to help capture the cyclic nature of SRL theory models, including Bayes networks, Directed Acyclic Graphs (DAG), Recurrent Neural Networks (RNN) and related time series models. 
Bayes and DAGs failed to meet suitable criteria due to the nature of SRL theory models being undirected, and that these technologies are predominantly focused on unidirectional weights~\cite{mumfordBayesianNetworksFMRI2014}. RNN was also found to not be suitable for this study as a significant amount of data was required for reliable training, and this data set had only a few hundred rows of reduced data.  RNN may be a viable consideration in the future, though many studies in this domain may not have a large enough volume of data~\cite{jarvelaCapturingDynamicCyclical2019}.

\subsection{Baseline Model}
\begin{figure*}[b]
\includegraphics[width=\textwidth]{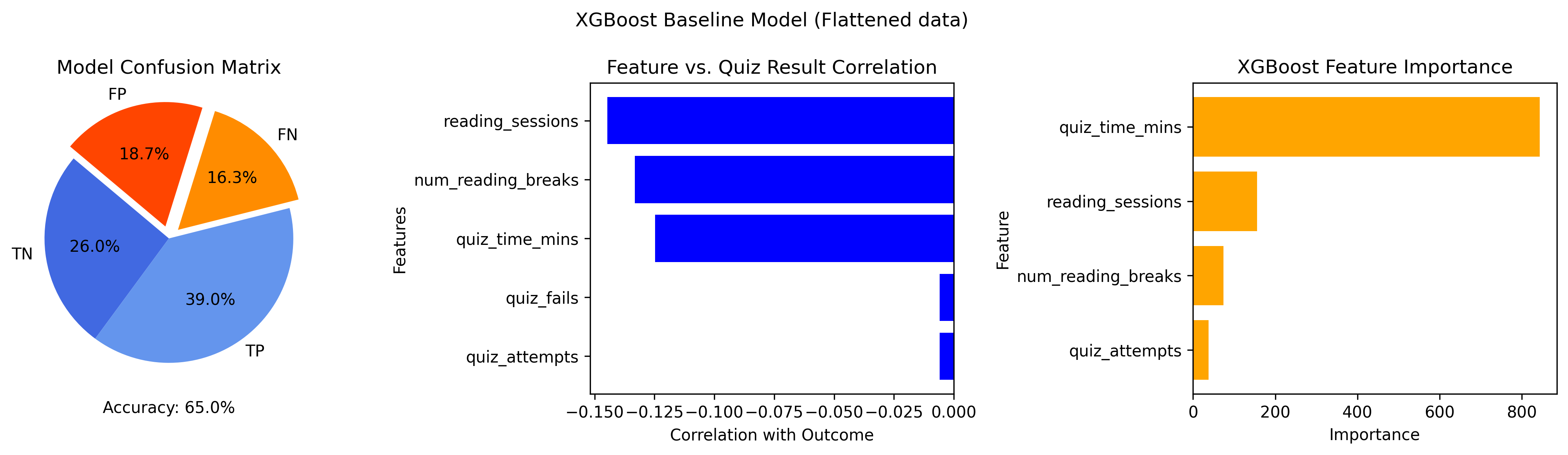}
\caption{Baseline XGBoost Results\label{fig1}}
\end{figure*}

After successful recreation of the core results of this study, additional features and modelling were done to analyse the features for their predictive power for academic quiz outcomes.

Common data cleaning process for many observation-based machine learning model pipelines is to calculate features and reduce them to a single row per-observation wide format~\cite{philuekMachineLearningTechniques2022}. This preparation works for most use cases, though in the case of trying to fit e-learning data into a theoretical self-regulated learning model, this methodology does not intuitively allow for cyclical, undirected  behaviours that we desire to model for prediction.

Since suitable models are not readily available for transforming and training models with a cyclical nature, our next step was to fall back to a trusted tree based model for analysis, and later engineer SRL focused features that describe the reinforcement behaviours.

As a baseline, the first model used only features from the original study and a pass/fail for each quiz as the target outcome; these features are displayed in \Cref{table1}.

While the study was primarily designed to cluster student reading behaviours, we see from the model results with XGBoost, displayed in \Cref{fig1}, that these features are not very strong in predicting a pass/fail outcome.  The accuracy for prediction was approximately 65\%.

\subsection{Engineering Cyclic Reinforcement Features}


\begin{wrapfigure}{R}{0.51\textwidth}
\centering
\includegraphics[width=0.5\textwidth]{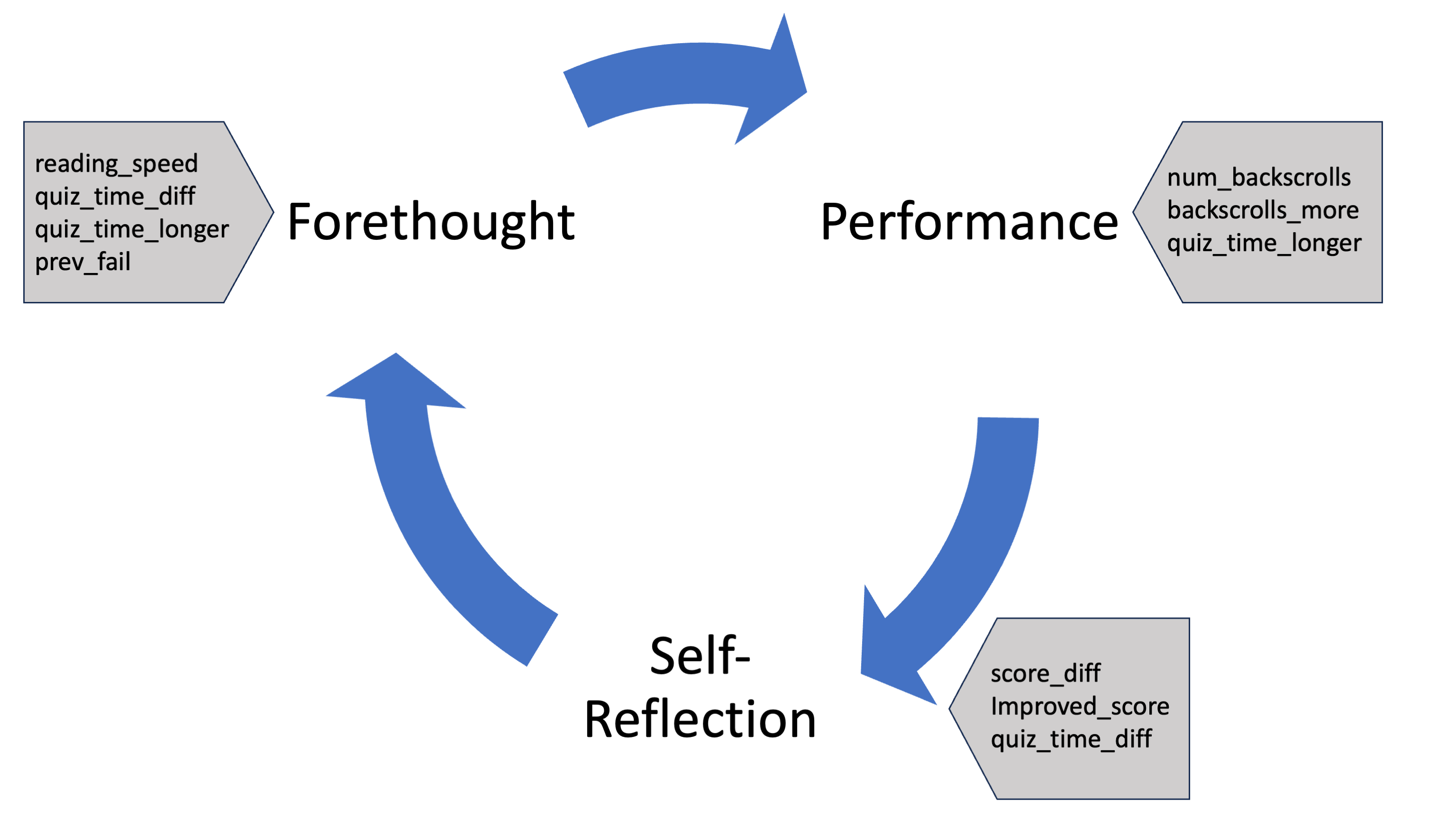}
\caption{Continuous Iterations of Self Regulated Learning Phases with Features Engineered to be Used in Subsequent Phases and Reinforcement Cycles\label{fig2}}
\end{wrapfigure}

\begin{figure*}[b]
\includegraphics[width=\textwidth]{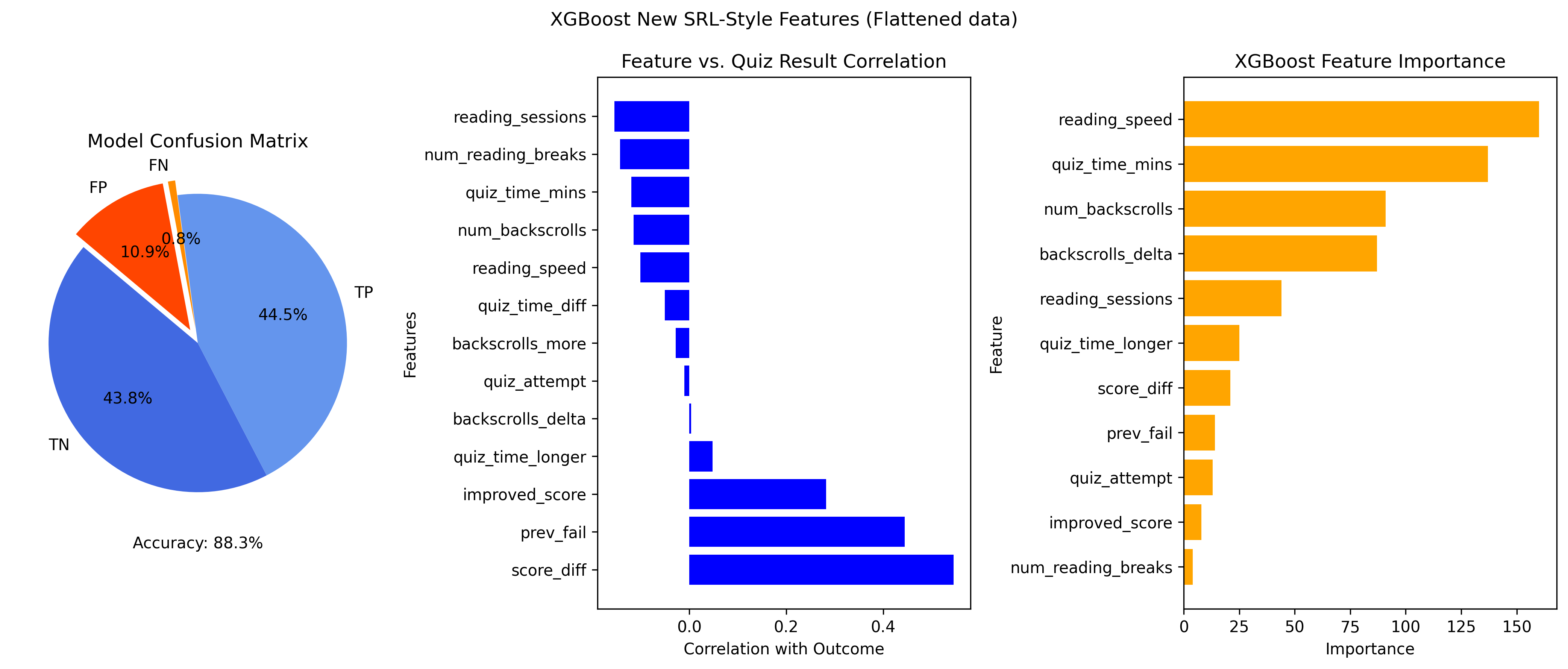}
\caption{SRL Enhanced XGBoost Results \label{fig3}}
\end{figure*}

SRL related features were then engineered from the same dataset. The hypothesis is that, if there is knowledge of previous attempt insight, behavioural attributes should significantly enhance the ability to predict students’ academic outcome if they exhibit observable SRL activities. The features are described below and displayed in the context of SRL theory in \Cref{fig2}.


\textbf{\textit{num\_backscrolls, backscrolls\_delta, backscrolls\_more}} : Features that count the number of backward scrolling actions while reading, and compare the count with previous quiz attempts.  These features relate to the performance phase, where students are recognizing the need to re-read, or scroll back for context, while reading new content.

\textbf{\textit{reading\_speed}} : An approximate metric was calculated based on how many page objects a user scrolled through in a given period of time. Faster reading may be a reflection phase behaviour, or a subsequent reading session may suggest execution phase regulation behaviour.

\textbf{\textit{prev\_fail, score\_diff}} and \textbf{\textit{improved\_score}}: Features that identify if the previous quiz attempt resulted in a failure, and the difference in the previous score.  These features help identify changes in behaviour as a result of self-reflection and may affect strategy selection in forethought planning.

\textbf{\textit{quiz\_time\_diff}} and \textbf{\textit{quiz\_time\_longer}} : The calculated difference between the current quiz duration and the previous quiz attempt.  A difference in time spent may indicate a change in the student’s performance regulation.

These features and explanations help to understand the motivation for their measurement.  Since we cannot yet measure the reinforcement effect internally from core metrics, we may simulate this by picking individual metrics that come from observed results from previous SRL studies.

When applying these features and using the same machine learning techniques, we see a notable difference in predictive accuracy as shown in \Cref{fig3}.

The results and analysis of this experiment confirm that SRL designed features significantly help catch the reinforcement signal of students' SRL activities, especially with features like \textit{score\_diff} and \textit{prev\_fail}.  While the baseline model from the previous study yielded only a 65\% accuracy rate, and substantial false positive and false positive results, the new model with SRL features increases performance to higher than 88\%, with a very low number of false negatives.

\section{Limitations and Further Research}
This case study is part of ongoing research into methods for cyclical modelling of SRL theory models.  As a result, limitations of these results include: (i) inclusion of only one study's existing peer-reviewed data, (ii) evaluation of only one model type (XGBoost), and (iii) only simple analysis of feature interdependence.  Our findings are being presented as work-in-progress, where the results support the value of further research into methods for cyclical modelling.

Further research will focus on separate models for each SRL phase data set, which enables targeting variables from alternate phases for prediction and explaining causal effects of SRL behaviors.  For example: Performance and Self-Reflection behaviors may help to predict subsequent forethought actions, just as Forethought behaviors may help predict upcoming performance behaviors.  By implementing separate models per phase, cross-phase analysis insights may help to understand the reinforcement relationship between phases, better map to SRL theory models, and provide improved SRL causal understanding when using trace data from e-Learning contexts.

\section{Conclusion}

Our results show that the addition of SRL oriented features significantly improves the prediction of XGBoost models with predicting quiz pass/fail outcomes.
The source study’s conditions and data processing were reproduced, and the baseline model yielded a 65\% accuracy rate.  After the SRL features were engineered with the same input data, the accuracy increased to 88\%. In addition, several of the new features were shown to be correlated with the target outcome, and also appeared in the top list of important features from the model analysis.

Results have the potential to be utilized as insights or to build scaffolding tools for students by post-analysis of class performance, and subpopulation analysis of behaviours associated with students that pass and fail using their history of SRL activities.  This can further be used to make predictive and recommender models that evaluate a student’s existing quiz history and current reading patterns to measure the likelihood of the student to pass or fail the upcoming quiz.  This prediction and an analysis of feature impact (using methods such as Shapley values) may further be used to make scaffold recommendations~\cite{chenAlgorithmsEstimateShapley2023}.

The results of this analysis emphasize the value and need for further research into modelling techniques that capture the cyclic nature of SRL in e-Learning environments.  This would further enable better understanding of causal effects within e-learning systems with explain-ability backed by existing SRL research.  The mapping of activities within e-learning and trace data to SRL theory models may further aid in understanding critical SRL strategy planning and execution within e-learning and to measuring the impact of interventions and external scaffolds such as GenAI.

\bibliographystyle{unsrt}  
\bibliography{references}

\end{document}